\documentclass[twocolumn,epjc3]{svjour3}          

\RequirePackage[T1]{fontenc}

\smartqed  

\RequirePackage{graphicx}
\RequirePackage{mathptmx}      
\RequirePackage{amsmath}
\RequirePackage{amssymb}
\RequirePackage{flushend}
\RequirePackage[numbers,sort&compress]{natbib}
\RequirePackage[colorlinks,citecolor=blue,urlcolor=blue,linkcolor=blue]{hyperref}
\RequirePackage{indentfirst}
\usepackage[todonotes={textsize=tiny}]{changes}
\definechangesauthor[name={Marcos}, color=red]{Marcos}
\definechangesauthor[name={Dedin}, color=purple]{Dedin}

\RequirePackage{changes}
\definechangesauthor[name={Marcos}, color=red]{M}
\definechangesauthor[name={Pedro}, color=blue]{P}

\journalname{Eur. Phys. J. C}

\begin{document}

\title{SN1987A neutrino burst: limits on flavor conversion}


\author{Pedro Dedin Neto\thanksref{e1,addr1,addr2}
        \and
        Marcos V. dos Santos\thanksref{e2,addr1}
        \and
        Pedro Cunha de Holanda\thanksref{e3,addr1}
        \and
        Ernesto Kemp\thanksref{e4,addr1}
}

\thankstext{e1}{e-mail: dedin@ifi.unicamp.br; orcid: 0000-0002-1831-3801 (corresponding author)}
\thankstext{e2}{e-mail: mvsantos@ifi.unicamp.br; orcid: 0000-0001-5247-744X}
\thankstext{e3}{e-mail: holanda@ifi.unicamp.br; orcid: 0000-0001-9852-8900}
\thankstext{e4}{e-mail: kemp@ifi.unicamp.br; orcid: 0000-0001-5311-1300}

\institute{Instituto de Física Gleb Wataghin, UNICAMP, Rua Sérgio Buarque de Holanda 777, Campinas-SP, Brazil\label{addr1}\and
Niels Bohr International Academy \& DARK, Niels Bohr Institute, University of Copenhagen, Blegdamsvej 17, 2100 Copenhagen, Denmark\label{addr2}
}

\date{}

\maketitle

\begin{abstract}
In this paper, we revisit the SN1987A neutrino data to see its constraints on flavor conversion. We are motivated by the fact that most works that analyze this data consider a specific conversion mechanism, such as the MSW (Mikheyev–Smirnov–Wolfenstein) effect, although flavor conversion is still an open question in supernovae due to the presence of neutrino-neutrino interactions. In our analysis, instead of considering a specific conversion mechanism, we let the electron antineutrino survival probability $P_{\overline{e}\overline{e}}$ be a free parameter. We fit the data from Kamiokande-II, Baksan, and IMB detected spectrum with two classes of models: time-integrated and time-dependent. For the time-integrated model, it is not possible to put limits above $1\sigma$ (68\% confidence level) on the survival probability. The same happens for the time-dependent model when cooling is the only mechanism of antineutrino emission. However, for models considering an accretion phase, $P_{\overline{e}\overline{e}}\sim0$ is strongly rejected, showing a preference for the existence of an accretion component in the detected antineutrino flux, and a preference for normal mass ordering when only the MSW is present.
\end{abstract}



\section{Introduction}
\label{sec:Introduction}
The detection of antineutrinos coming from the SN1987A supernova, the first and only detection of supernova neutrinos up to this date, was a big event for particle and astrophysics. The events were observed by the underground neutrino experiments Kamiokande-II (KII) \cite{Hirata:1987hu,Hirata:1988ad}, IMB 
\cite{Bionta:1987qt,Bratton:1988ww} and Baksan \cite{Alekseev:1988gp}. Since then, many works were produced to analyze and understand this data \cite{Loredo:2001rx, Pagliaroli:2008ur,Lunardini:2000sw,Lunardini:2004bj,Lunardini:2005jf,dos2022understanding}, which gave us information to put bound in supernova models and neutrino properties. However, some conditions used in previous works do not fit well in the picture that we have today. In this context, this paper is intended to be complementary to \cite{Loredo:2001rx, Pagliaroli:2008ur}.

One of the main questions regarding supernova neutrinos today is the flavor conversion mechanism. It is expected for the supernova neutrinos to suffer MSW conversion \cite{wolfenstein1978neutrino,mikheev1985resonance,Dighe:1999bi} and a substantial number of works were done considering this as the only conversion mechanism in action, including the ones that analyze the SN1987A data \cite{Loredo:2001rx,Pagliaroli:2008ur}. However, today it is expected that neutrino-neutrino interactions (forward scattering) become relevant in a supernova environment leading the neutrinos to a non-linear collective evolution \cite{duan2010collective}. Due to the complications that emerge from this type of evolution, there is not a conclusive picture of neutrino conversion in the supernova environment.

 Nevertheless, given the equal amount of non-electron antineutrinos $\overline{\nu}_x=(\overline\nu_\mu,\overline\nu_\tau)$ emitted from the supernova, it is possible to write the flavor conversion in terms of only the electron antineutrino survival probability $P_{\overline{e}\overline{e}}$. Therefore, we treat this probability as a free parameter to see how SN1987A data can constrain it. Something similar was done by F. Vissani in \cite{Vissani2014-me}. However, it seems that the influence of the survival probability is analyzed only for the MSW normal hierarchy scenario ($P_{\overline{e}\overline{e}}=0.64$) against the no oscillation one ($P_{\overline{e}\overline{e}}=0$). Here we take a more complete analysis for $P_{\overline{e}\overline{e}}$, allowing it to range from 0 to 1.

In section \ref{sec:Neutrino_Signal} we describe our model for the detected event rate in each detector (KII,IMB, Baksan) based on two different neutrino emission models, the flavor conversion mechanism, and the detection properties. In section \ref{sec:Statistical_Analysis} we describe our statistical analysis of the SN1987A data. In section \ref{sec:Results_Dicussion} we show our results and discuss them, and finally, in section \ref{sec:Conclusion} we  present our conclusions.

\section{Model for the neutrino signal}
\label{sec:Neutrino_Signal}
In this section, we describe the model for the expected neutrino event rate in each of the detectors, which is used to fit the SN1987A data. First, we describe the two neutrino emission models considered in this paper: a time-dependent and a time-integrated. In sequence, we describe the flavor conversion in the flux, which depends only on $P_{\overline{e}\overline{e}}$, and, in the end, we discuss the detection features of this analysis. Given that the most relevant cross-section for the considered detectors is the IBD, we will restrict our model to the antineutrino sector ($\bar{\nu}_e, \bar{\nu}_\mu, \bar{\nu}_\tau$)

\subsection{\textbf{Neutrino Emission}}

Based on previous SN1987A neutrino data analysis \cite{Loredo:2001rx,Pagliaroli:2008ur,Lunardini:2000sw, Lunardini:2004bj, Lunardini:2005jf}, we use two distinct models for the neutrino emission: time-integrated and time-dependent ones.

\paragraph{\textbf{Time-dependent}}
Given that the neutrino emission evolves in time, a time-dependent model should be at least considered in data analysis. This approach can be found in the famous paper of Lamb and Loredo \cite{Loredo:2001rx} and some other works \cite{Pagliaroli:2008ur}. In this approach, the antineutrino emission can be divided into two phases: the accretion and cooling phases. Here we will follow the path of \cite{Loredo:2001rx, Pagliaroli:2008ur} and model each phase by its most relevant mechanism of emission.

In this case, the accretion phase can be modeled as a positron thermal flux with temperature $T_a$ incident in a neutron target, that composes the mass in accretion in the proto-neutron star. Therefore, as in \cite{Loredo:2001rx, Pagliaroli:2008ur}, we consider that only electron antineutrinos are emitted in this phase and the flux is given by:
\begin{equation}
        \phi^0_{a, \bar{\nu}_{e}}(E_\nu,t) =\frac{8 \pi c}{(hc)^3} \left [ N_n(t) \sigma_{e^{+}n}(E_\nu) g_{e^+} (E_{e+},T_a) \right ],
\end{equation}
with
\begin{eqnarray}
     &&N(t)=\frac{Y_n}{m_n}\times M_a \times \frac{j_k(t)}{1+t/0.5s},   \nonumber \\
     &&g_{e^+} (E_{e+},T_a)=\frac{E_{e+}^2}{1+exp\left[E_{e+}/T_a\right]},
\end{eqnarray}
where $N_n(t)$ is the number of neutrons as a function of the time,  $\sigma_{e^{+}n}(E_\nu)$ the positron-neutron cross-section, and \linebreak $g_{e^+}(E_{e+},T_a)$ the thermal distribution of positrons with energy $E_{e+}$ in a temperature $T_a$. The number of neutrons is given by the initial accreting mass $M_a$ with a fraction of neutrons $Y_n$, and its time behavior is given  by the factor $j_k(t)=exp\left [ - \left ( t/\tau_a \right)^k \right]$, with  $\tau_a$ being the characteristic time of the accretion phase and the parameter $k=2$ following the parametrization in \cite{Pagliaroli:2008ur}\footnote{In \cite{Loredo:2001rx} it is used $k=10$, however, as discussed in \cite{Pagliaroli:2008ur} $k=2$ adjust better to supernova simulations.}. The denominator $1+t/0.5s$, as in \cite{Loredo:2001rx, Pagliaroli:2008ur}, is used to mimic the behavior from supernova simulations, where we have a constant flux within the first $0.5 \, s$ followed by a fast decrease.

The cooling phase, which is dominated by neutrinos and antineutrinos of all flavors emitted by the cooling neutron star, is modeled by a thermal distribution of fermions with temperature $T_c(t)$, with characteristic time  $\tau_c$, emitted from a sphere with fixed radius $R_c$ and is given by
 \begin{equation}
    \phi^0_{c,\bar{\nu}_{\alpha}}(E,t) = \frac{\pi c}{(hc)^3} 4\pi R_c^2 \frac{E^2}{1 + \exp[E/T_c(t)]},
\end{equation}
with the cooling temperature being a function of time
\begin{equation}
    T_c(t) = T_{c,\bar{\nu}_{\alpha}} \exp\left[ - t/\left ( 4\tau_c \right) \right].
\end{equation}

As already pointed out, different from the accretion component, the cooling one is composed of antineutrinos of all flavors. However, the non-electron antineutrinos $\overline{\nu}_x$ are emitted from deeper regions in the supernova, which can be effectively implemented by considering that they are emitted with higher initial temperatures $T_{c,\bar{\nu}_x}$. In fact, during the rest of the paper, we will talk about the ratio between the flavors temperatures $\tau=T_{\bar{\nu}_x}/T_{\bar{\nu}_e}$.

To combine the fluxes of both phases of emission, we follow \cite{Pagliaroli:2008ur}  where the cooling phase starts after the accretion one. As argued in the cited work, if the accretion and cooling phases were contemporaneous the first seconds would be composed of two different spectra, given the different temperatures of each of these phases. As numerical simulations of supernovae do not show this feature, we assume that the different emission phases are separated in time. We do this using the following parameterization:
\begin{equation}
    \phi^{0}_{\bar{\nu}} (t)=  \phi^0_{a}(t)+ (1-j_k(t))\phi^0_{c}(t-\tau_a),
\end{equation}
where the accretion flux is only composed of electrons antineutrinos $\phi^0_{a, \bar{\nu}_{e}}$, while the cooling flux contains an electronic $\phi^0_{c,\bar{\nu}_{e}}$  and non-electronic component $\phi^0_{c,\bar{\nu}_{x}}$.

\paragraph{\textbf{Time-integrated}} In this model, we consider that the time-integrated flux can be described by the following pinched spectrum \cite{Keil:2002in}:
\begin{eqnarray}
    \phi^0_{\beta}(E)&=&\frac{L_\beta}{E_{0\beta}}\frac{1}{(\alpha_\beta+1)^{-(\alpha_\beta+1)} \Gamma(\alpha_\beta+1)E_{0\beta}} \nonumber \\
    &&\times \left ( \frac{E}{E_0}\right)^{\alpha_\beta} e^{-(\alpha_\beta+1)E/E_{0\beta}},
\end{eqnarray}
where, for a specific neutrino flavor $\beta$, $L_\beta$ is the total energy (time-integrated luminosity), $E_{0\beta}$ the mean energy, and $\alpha_\beta$ the pinching parameter. We are mainly motivated to use this model due to a collection of works that only use the energy information from the SN1987A \cite{Lunardini:2000sw, Lunardini:2004bj, Lunardini:2005jf}. Although the time data could bring new information, it is interesting to check if the energy  alone can say something about the flavor conversion.

\subsection{\textbf{Flavor Conversion}}
From  emission until detection, the neutrino may suffer flavor conversion. It is still an open question for supernova neutrinos which is the complete mechanism of flavor conversion, given the complications that arise with neutrino-neutrino interactions. However, due to unitarity and the equal initial flux of non-electron antineutrinos $\phi^0_{\overline{\nu}_\mu}=\phi^0_{\overline{\nu}_\tau}=\phi^0_{\overline{\nu}_x}$, the equations for flavor conversion can be simplified so that it will only depend on the electron antineutrino survival probability $P_{\overline{e}\overline{e}}$ and initial fluxes \cite{Kuo:1989qe}, such that
\begin{subequations}
\begin{equation}
     \phi_{\overline{\nu}_e}= \phi^0_{\overline{\nu}_e} - (1-P_{\overline{e}\overline{e}})(\phi^0_{\overline{\nu}_e} - \phi^0_{\overline{\nu}_x}),
\label{eq:phie}
\end{equation}
\begin{equation}
    2 \phi_{\overline{\nu}_x} = 2 \phi^0_{\overline{\nu}_x} + (1-P_{\overline{e}\overline{e}})(\phi^0_{\overline{\nu}_e} - \phi^0_{\overline{\nu}_x}).
\label{eq:phix}
\end{equation}
\end{subequations}

Therefore, we can explore the survival probability $P_{\overline{e}\overline{e}}$ as a free parameter representing the flavor conversion occurring during the neutrino propagation. In this paper, we want to see how strong the SN1987A data can constrain $P_{\overline{e}\overline{e}}$ in the fitted models, given that the flavor conversion mechanism is still an open question in a supernova environment. Although this probability may be time and/or energy-dependent, we will consider it independent of these variables, given that we do not want to use a specific model.

We will also consider the MSW-only conversion scenario in order to compare it to our free $P_{\overline{e}\overline{e}}$ model. In this scenario, the electron antineutrino is created as a $\bar{\nu}_1$ for normal mass hierarchy (NH) and $\bar{\nu}_3$ for inverted mass hierarchy (IH). Therefore, the survival probability  for each mass ordering can be written as follows \cite{pagliaroli2007first}
\begin{subequations}
\begin{equation}
 P_{\overline{e}\overline{e}}^{\text{NH}} = U_{e1}^2= \cos\theta_{12}^2 \cos\theta_{13}^2,
\end{equation}
\begin{equation}
P_{\overline{e}\overline{e}}^{\text{IH}} = U_{e3}^2 = \sin\theta_{13}^2,
\end{equation}
\end{subequations}
where we have considered an adiabatic evolution, with a flipping probability equal to zero at the high and low-density resonances. The vacuum mixing parameters are taken from the update values published for the global fit analysis in \cite{Gonzalez-Garcia:2021dve}.

Although this energy dependence of $P_{\overline{e}\overline{e}}$ is negligible in the standard MSW effect, other possible effects associated with collective effects, such as spectral split among different neutrino flavors lead to a strong energy dependency, changing drastically this scenario \cite{duan2010collective}. However, given the unknowns associated with such collective effects nowadays, we limit our analysis to consider a $P_{\overline{e}\overline{e}}$ that is uniform in energy, leaving the spectral split analysis for a future work.

\subsection{\textbf{Detection}}

In the case of the SN1987A, we have data from three detectors: Kamiokande-II, IMB, and Baksan. In all of them, the dominant channel for electron antineutrino detection is the Inverse Beta-decay (IBD), which is the only one that we will consider. Therefore, the event rate $R^{\text{IBD}}_{\bar{\nu}_e}$ as a function of the positron measured energy $E_{e^+}$, the angle between the incoming neutrino and the scattered positron $\theta$ and time (for the time-dependent model) can be calculated as follows
\begin{eqnarray}
\label{eq:event_rate}
    R^{\text{IBD}}_{\bar{\nu}_e} (E_{e^+},t,\cos\theta)&=& N_p \times  \phi_{\bar{\nu}_e}(E_\nu, t) \nonumber \\
    &&\times \frac{d\sigma^{\text{IBD}}_{\bar{\nu}_e}}{d\cos\theta}(E_\nu)\times \eta^d(E_{e^+}),
\end{eqnarray}
where $N_p$ is the number of free protons, $\phi_{\bar{\nu}_e}(E_\nu, t)$ the electron antineutrino flux at the detector, $d \sigma^{\text{IBD}}_{\bar{\nu}_e}(E_\nu)/d\cos\theta$ the differential cross-section for IBD, and $\eta^d(E_{e^+})$ the detector intrinsic efficiency. For the IBD, the incoming neutrino energy $E_\nu$ is related to the created positron energy by $E_{e^+} \approx E_\nu-1.293 MeV$, due to the mass difference between the initial proton and the final neutron. The energy threshold for the IBD is $E_{\bar{\nu}}^{th} =1.806$ MeV \cite{Giunti_Carlo2007-lv}.

\subsection{Efficiency}
\label{sec:Efficiency}

As pointed out by \cite{Vissani2014-me}, when calculating the differential event rate in equation \ref{eq:event_rate}, one should use the detector intrinsic efficiency $\eta^d(E_{e^+})$. However, when integrating the event rate to get the total number of detected events, one should account for the threshold energy considered when selecting the events. This is achieved by multiplying the intrinsic efficiency by a function $g(E_{e^+},E_{min})$  resulting in a total efficiency 
\begin{subequations}
\begin{equation}
\label{eq:total_eff}
    \epsilon (E_{e^+},E_{min}) = \eta^d(E_{e^+})\times g(E_{e^+},E_{min}),    
\end{equation}
\begin{equation}
\label{eq:g_func}
    g(E_{e^+},E_{min})= \frac{1+\text{Erf}\left[\frac{E_{e^+} -E_{min}}{\sqrt{2} \sigma(E_{e^+})} \right]}{2},  
\end{equation}
\end{subequations}
in which the error function $\text{Erf}$ accounts for the threshold energy $E_{min}$ and the uncertainty $\sigma(E_{e^+})$ on the energy. This distinction between intrinsic and total efficiency is relevant when talking about the ones reported by the experiments, which are total efficiencies accounting for the threshold energies used during the events selections. This distinction becomes even more relevant in the case of the Kamiokande-II when using the low-energy events (numbers 13-16 nad 6 in table \ref{tab:Kamiokande_data}) added a posteriori and which are below the energy threshold of 7.5 MeV used in the first published data. To incorporate these events in our analysis, we need to infer the intrinsic efficiency from the published total efficiency and extrapolate the last to lower energies in the case of Kamiokande-II. Following this reasoning, we adopt the same parametrization for the intrinsic efficiency as reported in \cite{Vissani2014-me}, with $E_{min}=$ 4.5 MeV for Kamiokande-II. Both total and intrinsic efficiencies used in this work are shown in figure  \ref{fig:Detec_Eff}.

\subsection{Uncertainties}
The uncertainties used in this work are experimental ones shown in tables \ref{tab:Kamiokande_data}, \ref{tab:IMB_data}, and \ref{tab:Bakasan_data}. Although we have the angle uncertainty, we will not consider it in our analysis, due to its non-significant impact on the likelihood, given that the considered cross-section (IBD) has a weak angular dependency. Also, as pointed out in \cite{Pagliaroli:2008ur}, the relative time between the events is measured with good precision so that we also ignore the time uncertainty. As for the energy uncertainty, in addition to reported values for the energy of the events, to implement it in the efficiency expressions, such as equation in \ref{eq:g_func}, we need to estimate the uncertainty for other values of energy. For this purpose, we adopt an uncertainty parametrization with a statistical component that goes with the square root of the measured energy $E_{e^+}$ and a systematic one that grows linear with the energy. as done in \cite{Vissani2014-me}:
\begin{equation}
\sigma(E_{e^+}) = \sigma_{stat} \left(\frac{E_{e^+}}{10 MeV}\right)^{1/2} + \sigma_{syst} \left(\frac{E_{e^+}}{10 MeV}\right)
\end{equation}
The values that we used for the coefficients are shown in table \ref{tab:Detectors} corresponding to the ones that best adjust the function to the reported uncertainties.

\subsection{Cross-section}

The exclusive interaction considered in the analysis was the inverse beta decay, given the high cross-section compared to other possible channels of KII, IMB, and Baksan. We adopted the differential cross section (in the scattering angle) calculated by Vogel and Beacom in \cite{vogel1999angular}.

\subsection{Off-set time}
Another thing that we have to be careful of is to not confuse the time of the first detected neutrino $t_1$ with the time $t_0=t=0$ which indicates the time that the first neutrino arrives at the detector, even if it was not detected. Not considering this may force that the first detected neutrino is originated from the initial accretion phase, which may not be the case. As we will discuss later, for the MSW conversion in the inverted mass hierarchy scenario (IH), the initial $\bar{\nu}_e$ flux contributes only to $2\%$ of the detected flux, which makes it probable that the first detected neutrino came from the cooling phase and then $t_1 \neq t_0$. To get around this problem, it is usual to introduce an offset time $t^d_{\text{\text{off}}}=t_1-t_0$ between the first detected neutrino and the time of arrival of the first neutrino, which may be different for each detector given that they do not have an equal absolute time.

\subsection{Background Modeling}
\label{sec:Background}
In a realistic approach, we have to consider that detected events may come from background sources. The background rate is considered to be constant over the time of exposure, and also uniform over space, i.e., it depends only on the positron energy of the event $B=B(E_i)=d^2N_{B}/dtdE$. The independence regarding the spatial position is an approximation, given that there is more background at the wall of the detector, due to the surrounding material. 

The background can be measured and it is published by the collaborations. As argued in \cite{Costantini2006-df}, there is no need to do a convolution of these measured background rates with a Gaussian uncertainty in the energy, as done in \cite{Loredo:2001rx}, given that the background curve adjusted to the data already accounts for the uncertainty in the measurement. Therefore, one only needs to take the background rate from the experimental curve without doing a posteriori uncertainty convolution, which would double count the uncertainty effect. In our case, we use the background rate from \cite{Vissani2014-me} for both Kamiokande-II and Baksan, whereas the background is irrelevant for the IMB detector. In the case of the Time-Integrated analysis, we have to integrate the background rate in time to get the event rate per energy $B=B(E_i)=dN_{B}/dE$. The integration has to be done on the time of exposure to the supernova signal, i.e., the data-taking duration ($\sim 30 s$).

\section{Statistical Analysis}
\label{sec:Statistical_Analysis}
For the statistical analysis, we use the method of maximum unbinned likelihood, due to the low number of events. Our expression for the likelihood is similar to the one adopted in \cite{Pagliaroli:2008ur}

\begin{eqnarray}\label{eq:likelihood}
    \mathcal{L} &=& e^{-f_d \int R(t) dt}\prod_{i=1}^N e^{R(t_i)\tau_d} \nonumber \\
    &&\times \left[\frac{B_i}{2}+\int R(t_i,E_{e,i},\cos\theta_i)\mathcal{L}_i(E_e) dE_e \right] .
\end{eqnarray}

Here we made implicitly the dependency of $\mathcal{L}$ in the parameters of our models. In this equation, $i$ is the index of each event, $R(t,E,\cos\theta)$ is the expected event rate from equation (\ref{eq:event_rate}), $R(t)$ the event rate integrated in the angle and energy, and $B$ the background rate\footnote{The factor of $1/2$ in the background rate term comes from its angular dependency in $\cos \theta$, which we consider to be uniform.} discussed in section \ref{sec:Background}. Here we differ from \cite{Pagliaroli:2008ur} in the definition of $R(t)$, in which we consider the total efficiency to calculate the event rate integrated in the energy, as discussed in section \ref{sec:Efficiency}. The integration in the positron energy $E_e$ is made considering a Gaussian distribution $\mathcal{L}_i(E_e)$ around the measured value $E_{e,i}$ with standard deviation given by the measurement uncertainty. As already discussed, we consider that the time and angle uncertainties are irrelevant. We also consider the dead time $\tau_d$ for each detector ($d=K,B,I$), where $f_d$ is the live-time fraction \cite{Pagliaroli:2008ur}. In the case of the time-independent model, we only have to consider a time integration in the event rate for the signal $R(t_i,E_{e,i},\cos\theta_i)$ and for the background $B(E_i)$.

To find the set of parameters that best adjusts our model to the data, we only have to maximize the likelihood $  \mathcal{L}$ or minimize $-2\log(\mathcal{L})$. The last one is useful because it transforms multiplication into a sum and has a straightforward connection to confidence intervals. Given that we have a set of parameters $\vec{\theta}$, taking their the best-fit $\hat{\vec{\theta}}$ we can define the likelihood ratio as follows.
\begin{equation}
 \lambda(\vec{\theta}) \equiv \mathcal{L}(\vec{\theta})/\mathcal{L}(\hat{\vec{\theta}}) 
\end{equation}
so that $-2 \log \lambda(\vec{\theta})$ follows a $\chi^2$ distribution in the asymptotic limit of large samples $N\rightarrow\infty$, with $m$ degrees of freedom representing the number of parameters not constrained to be in its best-fit value. With this procedure, we can estimate the best-fit values for the parameters and their confidence interval, given a confidence level. However, we have to note that our data is not a large sample so our confidence level is an approximation. In any case, in this paper, we consider that it is an acceptable approximation given the allowed region for the astrophysical parameters to be comparable to previous works \cite{Loredo:2001rx} that use other approaches to set the confidence levels, as we discuss in \ref{app:Compare_Other_Works}.

\section{Results and Discussion}
\label{sec:Results_Dicussion}
\subsection{\textbf{Time-dependent model}}\label{subsec:time-dependent-model}

 For the time-dependent model, following the references \cite{Loredo:2001rx,Pagliaroli:2008ur}, we consider two possible cases, one with just cooling emission and the other with an initial accretion phase. For the cooling component, we have four astrophysical parameters, the initial cooling temperature $T_c$, the time constant of the phase $\tau_c$, the radius of the neutrinosphere $R_c$, and the ratio between the initial temperatures of the electronic and non-electronic antineutrinos $\tau=T_{\bar{\nu}_x}/T_{\bar{\nu}_e}$. Previous works \cite{Pagliaroli:2008ur} fix this temperature ratio based on supernova simulations. Here, we check the impact of changing this ratio given that it has strong implications in how similar the initial spectra are, which reflects how well we can identify flavor conversion in the detected spectrum. Nevertheless, we limit ourselves to the range of temperature ratio expected from supernova simulations \cite{Keil:2002in}. When considering the accretion phase, we introduce three new astrophysical parameters: the initial accretion temperature $T_a$, the time constant of the phase $\tau_a$, and the accretion mass $M_a$. In addition to the astrophysical parameters, there is the offset time for each detector and the survival probability, resulting in a total of 8 parameters for the cooling model and 11 for the cooling plus accretion.

\begin{figure}
    \centering
    \includegraphics[width=\linewidth]{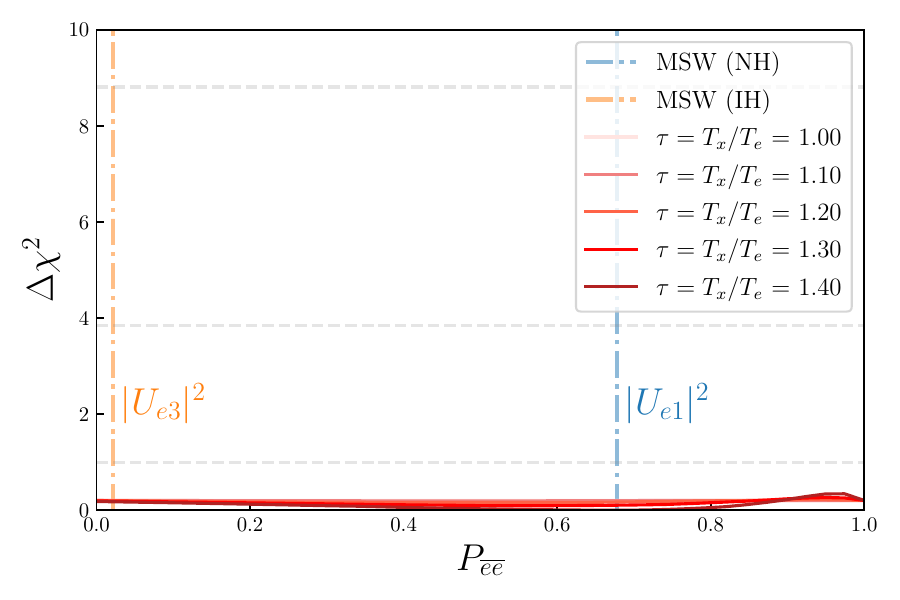}
    \caption{$P_{\overline{e}\overline{e}}$ likelihood ratio ($\Delta \chi^2 =-2\log\mathcal{L}/\mathcal{L}_{max}$) for the SN1987A data considering the time-dependent model with only the cooling component. The horizontal dashed lines correspond to $1$, $2$ and $3\sigma$ of C.L. Note that  minimum $\chi^2_{min}=-2\log\mathcal{L}_{max}$ is the one absolute regarding all the curves.}
    \label{fig:Pee_profile_time_dependent_Coo}
\end{figure}

\begin{figure}
    \centering
    \includegraphics[width=\linewidth]{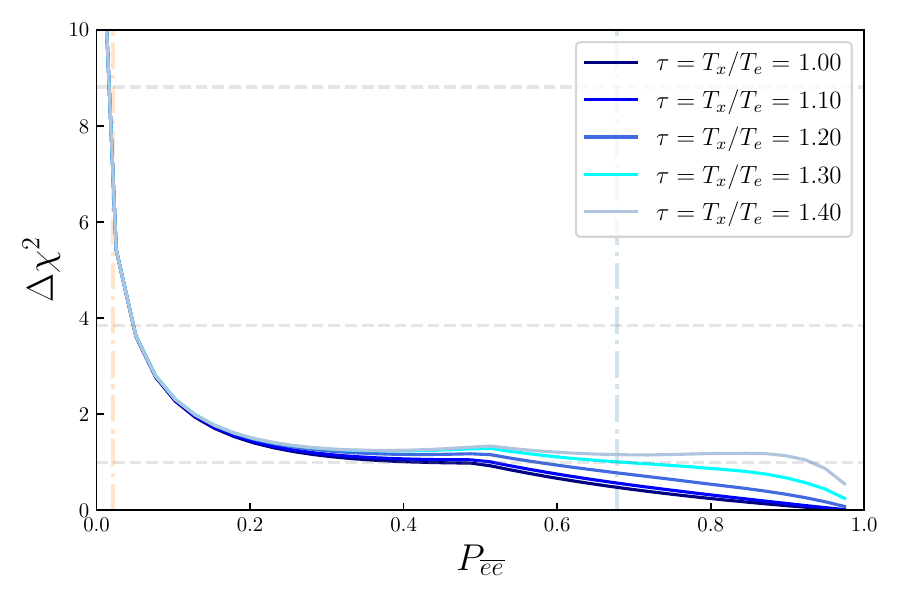}
    \caption{Same as Fig.~\ref{fig:Pee_profile_time_dependent_Coo} with two components: accretion and cooling. In this case, the two phases are considered to be separated in time. The horizontal dashed lines corresponds to $1$, $2$ and $3\sigma$ of C.L.}
    \label{fig:Pee_profile_time_dependent_Coo_Acc}
\end{figure}

To analyze how the SN1987A data can put limits on $P_{\overline{e}\overline{e}}$, we can do a marginal analysis, as described in section \ref{sec:Statistical_Analysis}. Figures \ref{fig:Pee_profile_time_dependent_Coo} and \ref{fig:Pee_profile_time_dependent_Coo_Acc} show the marginal plot of $P_{\overline{e}\overline{e}}$ for the models with only cooling component and for the one with cooling and accretion, respectively. For the model with just cooling, we can see that it is not possible to put limits on $P_{\overline{e}\overline{e}}$ up to the $1\sigma$ for $\tau$ values considered. This probably happens because both initial fluxes $\phi_{\overline{\nu}_e}^0$ and $\phi_{\overline{\nu}_x}^0$ come from the same mechanism, resulting in almost indistinguishable spectra, even allowing the temperatures to be different.

When we consider the accretion phase, we have a different scenario, where $P_{\overline{e}\overline{e}} \sim 0$ is strongly rejected, as we can see in Figure \ref{fig:Pee_profile_time_dependent_Coo_Acc}. This stronger constraint in $P_{\overline{e}\overline{e}}$ happens because in the accretion mechanism only electrons antineutrinos are emitted, making their initial flux $\phi^0_{\overline{\nu}_e}$ more distinguishable from the non-electronic one $\phi^0_{\overline{\nu}_x}$, which in turns facilitates the identification of flavor conversion. Given that, the excluded region of $P_{\overline{e}\overline{e}} \sim 0$ corresponds to the case where the detected flux is composed only by the initial $\phi^0_{\overline{\nu}_x}$, i.e., a flux with no accretion component. This shows us that the detected electron antineutrinos are better described by a flux with an accretion component coming from $\phi^0_{\overline{\nu}_e}$, as already found by \cite{Loredo:2001rx}. However, in \cite{Loredo:2001rx} they do not consider the role of flavor conversion, while here we can see that the existence of an accretion component has strong implications on the conversion mechanism. If we consider only the MSW effect with adiabatic propagation, this implies that the normal hierarchy scenario is favored over the inverted. Comparing them with the best-fit of free $P_{\overline{e}\overline{e}}$, the normal hierarchy scenario is not significantly rejected, while the inverted one is rejected by $\sim 3 \sigma$ of significance. 

\begin{figure}
    \centering
    \includegraphics[width=\linewidth]{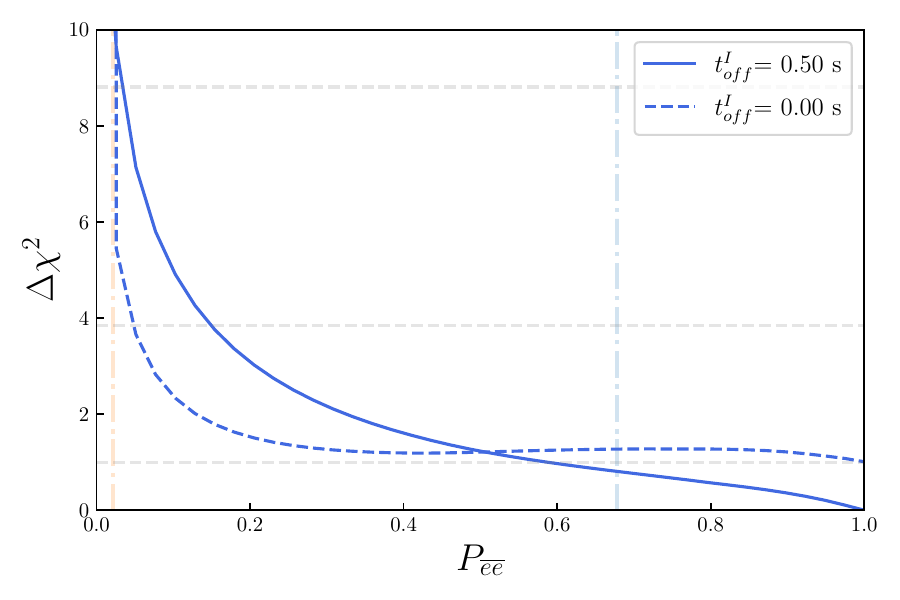}
    \caption{Same as Fig. \ref{fig:Pee_profile_time_dependent_Coo_Acc} but fixing $\tau=1.2$ for two different values of off-set time for the IMB data $t_{off}^I$}
    \label{fig:Pee_profile_time_dependent_Coo_Acc_toff_I}
\end{figure}

It is also possible to see in figure \ref{fig:Pee_profile_time_dependent_Coo_Acc} some kind of discrete transition to a lower $\Delta \chi^2$ at $P_{\overline{e}\overline{e}}\sim 0.5$. This happens because there is a preference for a non-zero off-set time in the IMB data, as can be seen in the best-fit value of $t_{off}^I$ in table \ref{tab:parameters_time-dependent} if the accretion component is strong enough (MSW-NH or free $P_{\overline{e}\overline{e}}$). However, if we go to lower values of $P_{\overline{e}\overline{e}}$, such as in the MSW-IH, it becomes preferable to describe some of the first events of IMB as coming from the cooling, i.e. $t_{off}^I=0$. This transition can be seen in figure \ref{fig:Pee_profile_time_dependent_Coo_Acc_toff_I} in which we plot the $\Delta\chi^2$ profile for $t_{off}^I=$ 0.5 and 0 s.

\begin{figure}
    \centering
    \includegraphics[width=\linewidth]{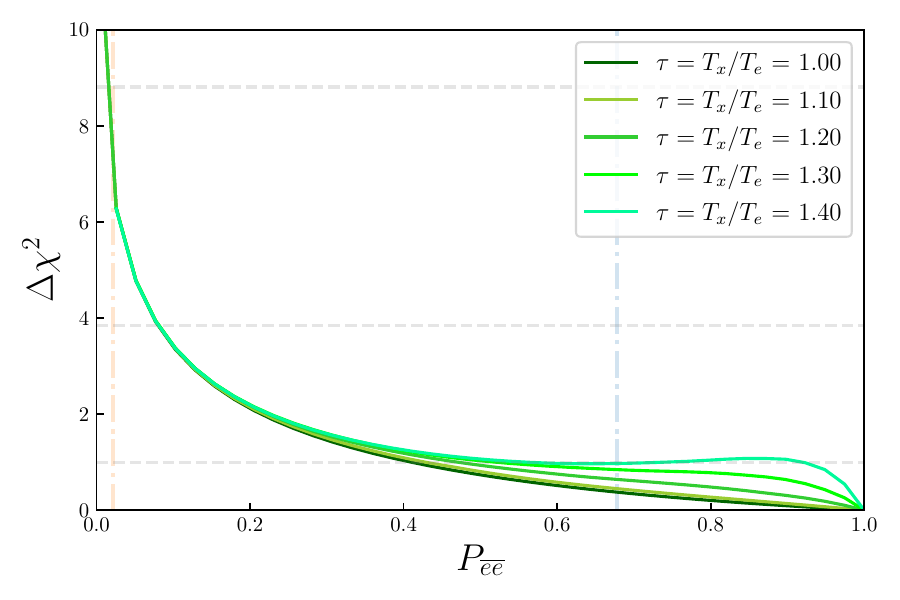}
    \caption{Same as Fig. \ref{fig:Pee_profile_time_dependent_Coo} with two components: accretion and cooling. In this case, the two phases are considered to be contemporaneous. The horizontal dashed lines corresponds to $1$, $2$ and $3\sigma$ of C.L.}
    \label{fig:Pee_profile_time_dependent_Coo_Acc_Cont}
\end{figure}

We have also tested the implications of considering the cooling and accretion components as contemporaneous. As argued by \cite{Pagliaroli:2008ur}, there is no evidence of a composed spectrum in supernova simulations, so the two mechanisms with different mean energies should occur at different times. However, from supernovae physics, we may expect that the PNS starts to cool down by neutrino emission soon after its formation, simultaneously with the accretion mechanism \cite{Olsen:2022pkn}. Therefore, we decide to test the implications of that hypothesis in our analysis. As we can see in Figure \ref{fig:Pee_profile_time_dependent_Coo_Acc_Cont} there is no significant modification on $P_{\overline{e}\overline{e}}$ limits. The only modification appears on the best-fit of $t_\text{off}^\text{IMB}$, which can be seen in \ref{app:Compare_Other_Works}.

\subsection{\textbf{Time-integrated model}}\label{subsec:time-integrated-model}

For the time-integrated model, we considered a Fermi-Dirac emission ($\alpha_{\overline{\nu}_e}=\alpha_{\overline{\nu}_x}=2.3$), a choice that does not have big impact in the fitting for $2.3<\alpha<4$ \footnote{By letting $\alpha_{\overline{\nu}_e}$ and $\alpha_{\overline{\nu}_x}$ run free in this interval, the variation of the likelihood ratio $\mathcal{L}/\mathcal{L}_{max}$ was not above 1$\sigma$ (C.L. $\approx 68\%$).}. We also consider a hierarchy for the mean energy $\overline{E}_{\overline{\nu}_x}>\overline{E}_{\overline{\nu}_e}$, which is physically motivated given that non-electron neutrinos interact less (lack of $\tau$ and $\mu$ leptons in the environment) and then escape from deeper regions in the supernova with higher temperatures. The best-fit values for the astrophysical parameters are shown in Table \ref{tab:parameters_time-integrated} considering the 3 different conversion scenarios. As we can see, there is a preference for a detected spectrum $\phi_{\overline{\nu}_e}$ to be composed mostly by the initial non-electron neutrino spectrum $\phi^0_{\overline{\nu}_x}$, given that there is basically no constraint for the total energy $\varepsilon_{\overline{\nu}_e}$, the same behavior was also found in \cite{Lunardini:2005jf}. Even in the MSW mechanism with inverted mass hierarchy, where the composition of $\phi^0_{\overline{\nu}_x}$ in the final flux is small ($P_{\overline{e}\overline{e}}\approx 2.18 \%$, the flavor conversion is compensated by a higher total energy $\varepsilon_{\overline{\nu}_x}$. This preference is a combination of the imposed energy hierarchy $\overline{E}_{\overline{\nu}_x}>\overline{E}_{\overline{\nu}_e}$ and the low detection efficiency for lower energies, where the low energy events can be as well described as coming from the background. However, we did not investigate this preference deeply\footnote{We only tested a scenario with relaxed bound conditions for the parameters. However, we obtained nonsensical values for the electron antineutrino total energy, such as $\varepsilon_{\overline{\nu}_e} \sim 10^{55} \text{ergs}$ for the inverted mass hierarchy.}. As we are interested in the flavor conversion parameter $P_{\overline{e}\overline{e}}$, we leave the \ref{app:Compare_Other_Works} to compare our marginal and contour plots with previous analyses to show the consistency of our method, at least regarding the astrophysical parameters.

\begin{figure}
    \centering
    \includegraphics[width=\linewidth]{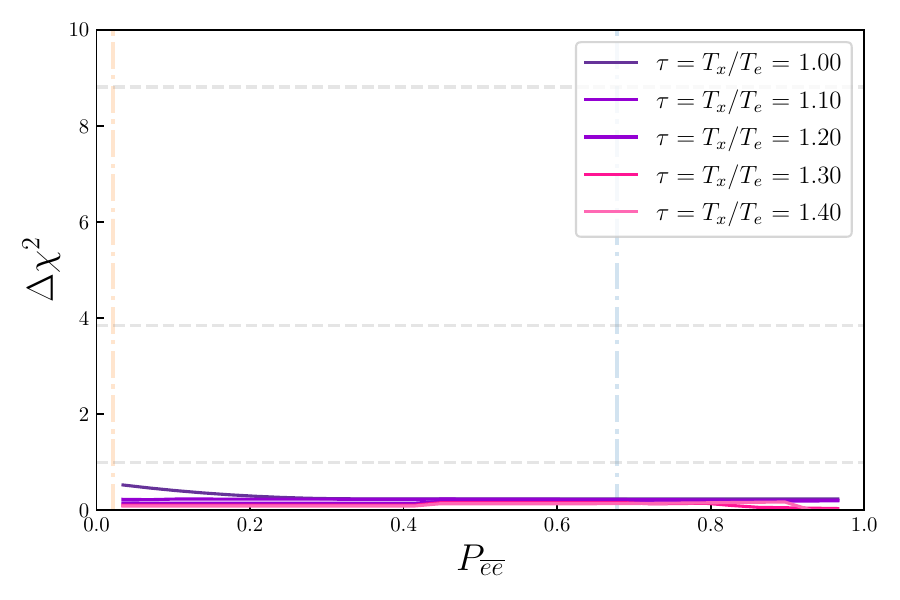}
    \caption{$P_{\overline{e}\overline{e}}$ likelihood ratio  for the SN1987A data considering the time-integrated model.} 
    \label{fig:Pee_profile_time_integrated}
\end{figure}

For the flavor conversion analysis, we again fix the initial temperature ratio (more precisely the mean energy ratio $\tau=\overline{E}_{\overline{\nu}_x}/\overline{E}_{\overline{\nu}_e} = T_{\overline{\nu}_x}/T_{\overline{\nu}_e}$) and let the other parameters run freely over the allowed range (Table \ref{tab:parameters_time-integrated}). Figure \ref{fig:Pee_profile_time_integrated} shows the marginal plot of $P_{\overline{e}\overline{e}}$ minimizing over the other model parameters. Again, there is no constraint on the survival probability above $68\%$ of confidence, even for spectra with higher mean energy differences such as $\tau=1.4$.

\subsection{Problems with fitting the data with some models}
\label{sec:numerical_problems}

In our numerical implementation, we found some difficulties in working with the two-component model (accretion + cooling). The main one is the existence of different local minima, which make the minimizer algorithm give different best fits depending on the initial conditions. To get around this problem, we  used two methods to find the global minimum. In the first method we fit this model multiple times ($\approx 1000$) fluctuating the initial conditions of parameters uniformly in the ranges shown in Table \ref{tab:parameters_time-dependent}, and taking the minimum value of $-2\log\mathcal{L}$ as the initial condition to find the global best-fit. The second method was based on using different minimizers (MINOS, scipy, simplex)\footnote{All of them implemented in the iminuit library \cite{iminuit}.} to see if this dependency on the initial conditions was algorithm dependent. In the end, we found that all the different minimizers obtained the same best fit given initial conditions around it, and in agreement with the first method. Given the concordance between the two methods and algorithms, we have confidence that the best fit obtained is the most probable one inside the allowed parameter space.

\section{Conclusion}
\label{sec:Conclusion}
In this paper, we have explored the role of flavor conversion in the SN1987A neutrino data, and how it can impose limits on the flavor conversion mechanism. We found that the time-integrated model, which uses only the energy information, could not put any limit on the electron antineutrino survival probability $P_{\overline{e}\overline{e}}$. The same happens for the time-dependent models that consider antineutrino emission only from the cooling mechanism. However, with  the existence of an accretion emission of electron antineutrinos, strong limits are imposed on low values of $P_{\overline{e}\overline{e}}$. This is impressive given the low statistics of the SN1987A neutrino data and it is in agreement with the previous work of Lamb and Loredo \cite{Loredo:2001rx} in which the data shows a strong preference for the existence of an accretion component. 

In previous works, such as \cite{pagliaroli2007first}, it was already pointed out that the inverted mass hierarchy was disfavored in MSW adiabatic scenario with a significance of 3$\sigma$ for some values of $\theta_{13}$, which was unknown at that time. Here we confirm this statement, as it can be seen from the figures \ref{fig:Pee_profile_time_dependent_Coo_Acc} and \ref{fig:Pee_profile_time_dependent_Coo_Acc_Cont}. Our improvement to their analysis was to use the current well-known neutrino vacuum mixing angles \cite{Gonzalez-Garcia:2021dve} and extend the analysis to the whole spectrum of possible values for the survival probability $P_{\overline{e}\overline{e}}$.

As we discussed, our analysis does not consider any time or energy dependency on $P_{\overline{e}\overline{e}}$, which may happen when we consider collective effects due to neutrino-neutrino forward scattering. We leave the study of time and energy dependency for a future paper. In any case, our results can still be used to constrain conversion models that result in a fixed value for $P_{\overline{e}\overline{e}}$.

\section*{Acknowledgments}
This work was supported by the Fundação de Amparo à Pesquisa do Estado de São Paulo (FAPESP) grants no. 2019/08956-2, no. 14/19164-6, and no. 2022/01568-0 and also financed in part by the Coordenação de Aperfeiçoamento de Pessoal de Nível Superior – Brasil (CAPES) – Finance Code 001. This version of the article has been accepted for publication, after peer review but is not the Version of Record and does not reflect post-acceptance improvements, or any corrections. The Version of Record is available online at: \url{http://dx.doi.org/10.1140/epjc/s10052-023-11597-6}.

\bibliography{mybibfile.bib}
\bibliographystyle{unsrt} 
\appendix

\section{Comparing results with other works}
\label{app:Compare_Other_Works}
Here we show our results for the astrophysical parameters fit in the format of marginalized profile and contour plots for each individual parameter and contour plots for some key combination of parameters.

\subsection{Time-Dependent}

\begin{figure}
    \centering
    \includegraphics[width=0.52\textwidth]{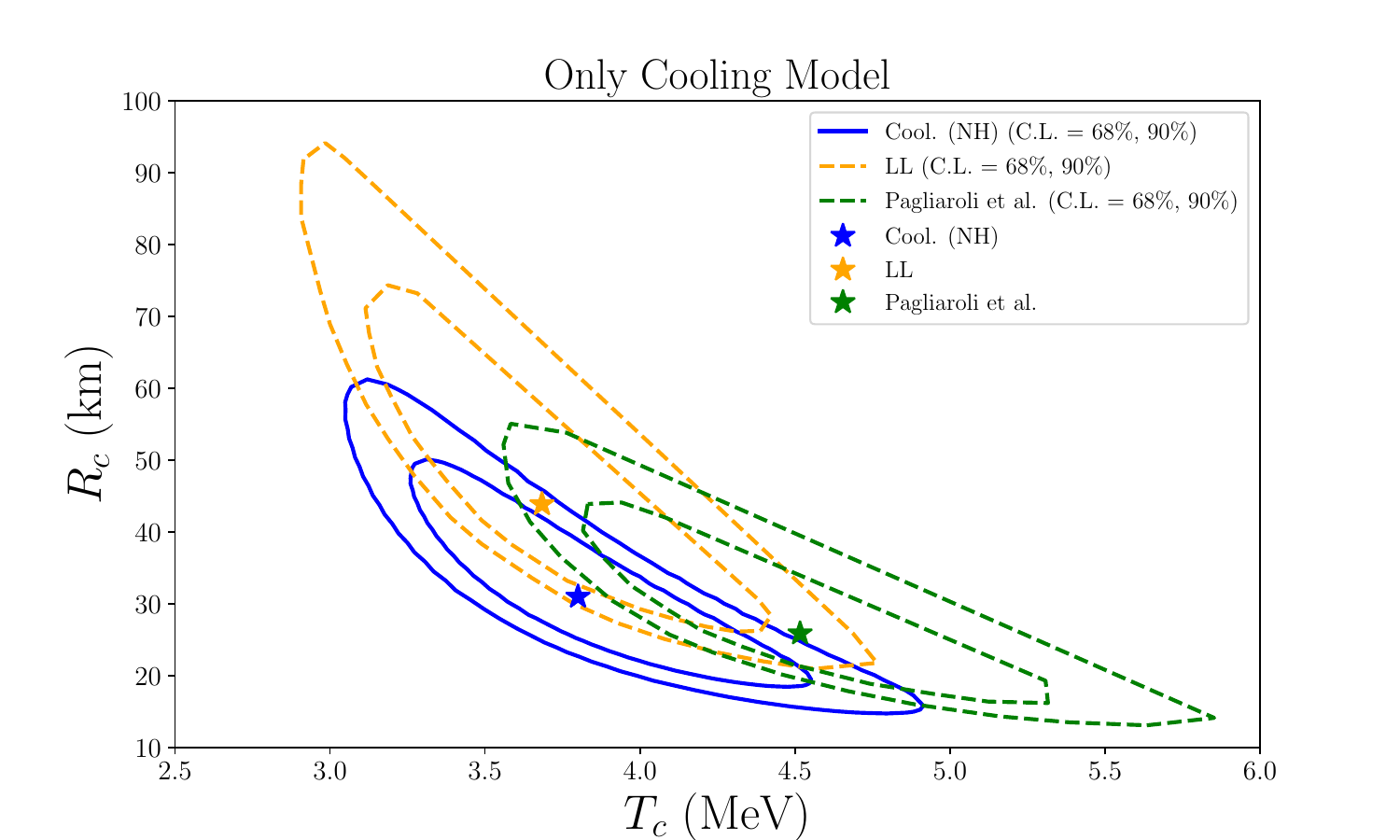}
    \caption{$T_{c,0}$ vs $R_c$ contour plots comparing our results with previous ones \cite{Loredo:2001rx,Pagliaroli:2008ur}.}
    \label{fig:Tc_Rc_contour}
\end{figure}

\begin{figure*}[h!]
    \centering
    \includegraphics[width=\linewidth]{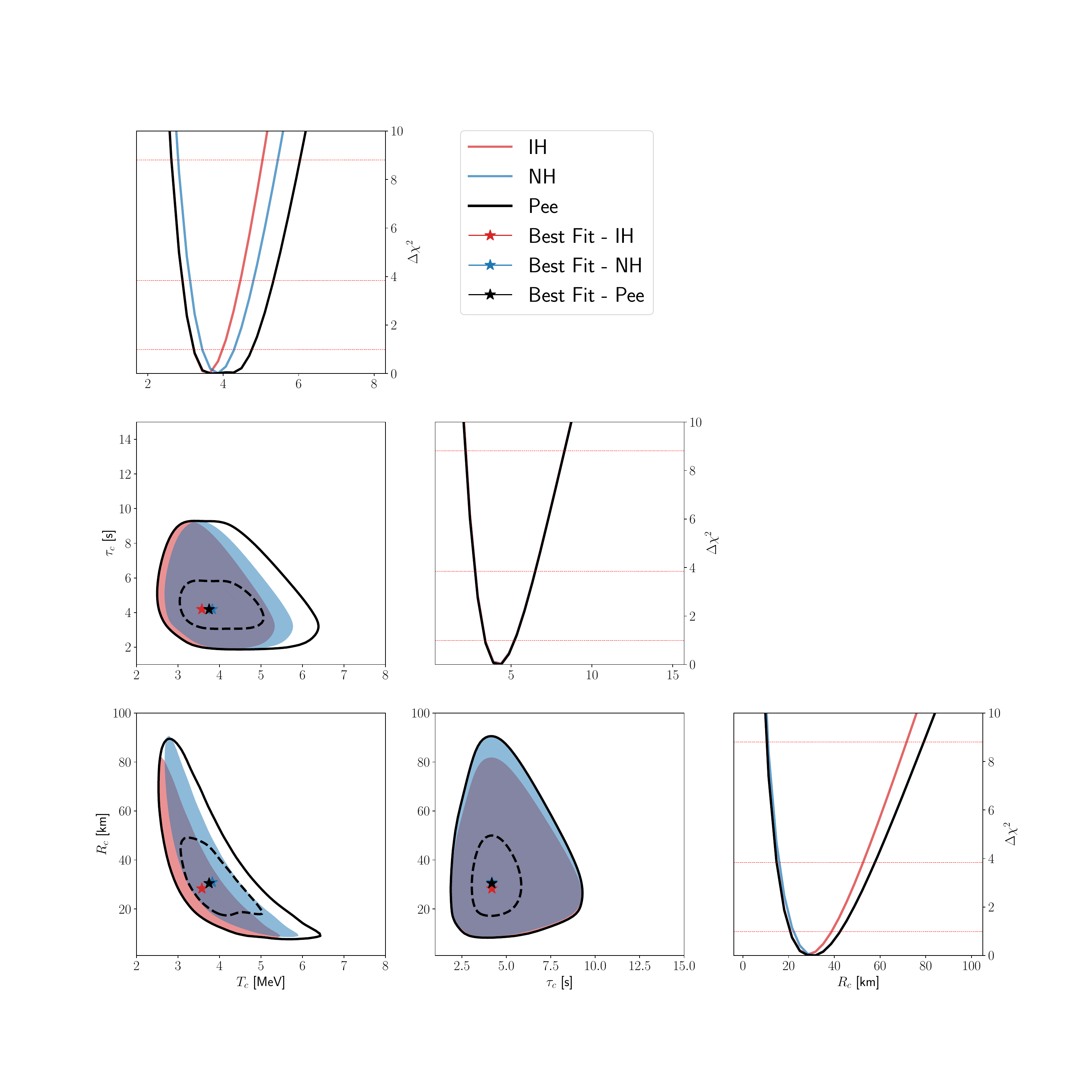}
      \caption{Marginal and contour plots for the astrophysical parameters $T_c,\tau_c,R_c$ for the only cooling model, keeping the detection off-set times $t^{\text{KII}}_{\text{off}},t^{\text{IMB}}_{\text{off}},t^{\text{Bak}}_{\text{off}}$ in their best-fit value. For the contour plots, we use color bands for the MSW-IH and MSW-NH scenarios and lines for the free $P_{\bar e \bar e}$, corresponding to confidence levels of 68\% (dashed) and 99.7\% (solid). Note that  minimum $\chi^2_{min}=-2\log\mathcal{L}_{max}$ is the absolute one among all the curves/conversion scenarios.}
      \label{fig:Param_Triangle_Profile_Time_Dependent_Coo}
\end{figure*}

\begin{figure*}[h!]
    \centering
    \includegraphics[width=\linewidth]{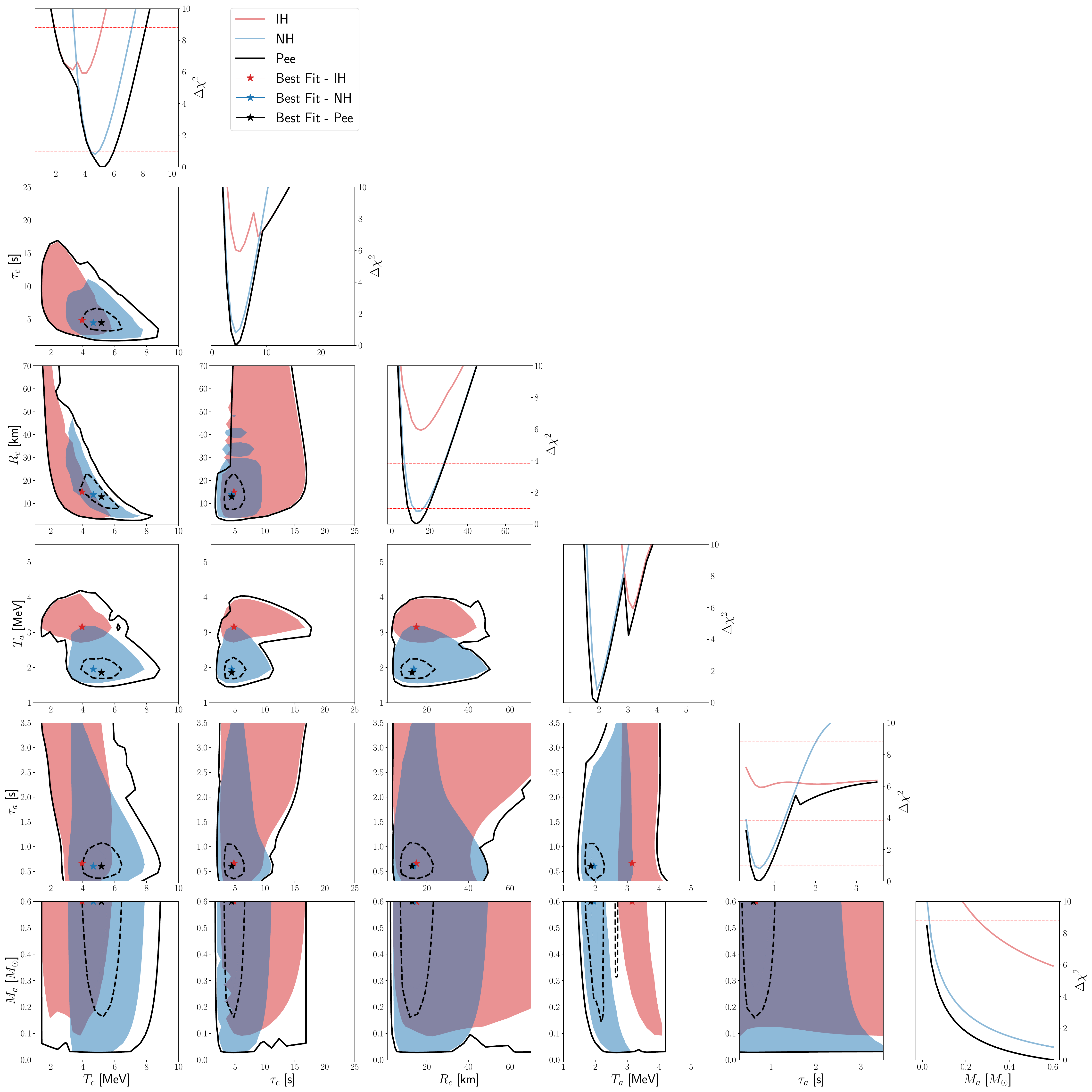}
      \caption{Same as figure \ref{fig:Param_Triangle_Profile_Time_Dependent_Coo}, but including the accretion component with parameters new parameters $T_c,\tau_c,R_c,T_a,\tau_a,M_a$.}
    \label{fig:Param_Triangle_Profile_Time_Dependent_Coo_Acc}
\end{figure*}

The results of the time-dependent analysis are very comparable to Loredo and Lamb \cite{Loredo:2001rx} and Pagliaroli et al. \cite{Pagliaroli:2008ur} work. Both authors also analyzed SN1987A data to respect to the same time-dependent model used here. In figure \ref{fig:Tc_Rc_contour}, we show the statistical limits on $T_c \times R_c$. Our bounds overlap with both works but it is not identical to them. We attribute this difference to our different implementation of the efficiencies, as discussed in section \ref{sec:Efficiency} and shown in Figure \ref{fig:Detec_Eff}, in addition to the use of updated neutrino mixing parameters.For a more complete view of our analysis and results, we also show $\Delta\chi^2$ profiles and contour plots of the astrophysical parameters for the model with only cooling (Figure \ref{fig:Param_Triangle_Profile_Time_Dependent_Coo}) and the one with cooling and accretion (Figure \ref{fig:Param_Triangle_Profile_Time_Dependent_Coo_Acc}), as well as the best values found and intervals used shown in Tables \ref{tab:parameters_time-dependent_coo} and \ref{tab:parameters_time-dependent}. It is possible to see in the plots that the profile for each parameter agrees with the obtained contour plots. Also, the conversion model with free $P_{\overline{e}\overline{e}}$ encompasses the MSW-IH and MSW-NH scenarios, as one would expect given that the latter are specific cases from the former, with $P_{\overline{e}\overline{e}}\approx 2.18 \%$ and $P_{\overline{e}\overline{e}}\approx 67.8 \%$ respectively.

\begin{table}[ht!]
\centering
\caption{Range and best-fit (BF) for all parameters in the time-dependent model Only Cooling. We show the best-fit for three flavor conversion scenarios: MSW with NH, MSW with IH, and a model-independent free $P_{\overline{e}\overline{e}}$.}
\label{tab:parameters_time-dependent_coo}
\begin{tabular}{ lllll } 
 \hline
Parameter & NH BF & IH BF & Free $P_{\overline{e}\overline{e}}$ BF & Range \\ 
 \hline
 $T_{0,c}$ [MeV]   & $3.8 ^{+0.5}_{-0.4}$  & $3.6 ^{+0.4}_{-0.4}$ &$3.7 ^{+1.0}_{-0.5}$ & 1-10 \\
 $\tau_c$ [s]      & $4.2 ^{+1.0}_{-0.8}$     & $4.2 ^{+1.0}_{-0.8}$ &$4.2 ^{+1.0}_{-0.8}$      & 1-40 \\
 $R_c$ [km]        & $31 ^{+12}_{-9}$        & $28 ^{+11}_{-8}$     &$30 ^{+12}_{-10}$      & 1-100\\
 
 $t^{\text{KII}}_{\text{off}}$ [s]   &$0.0 ^{+0.18}_{-0}$ &$0.0 ^{+0.18}_{-0}$ &$0.0 ^{+0.18}_{-0}$ & 0-6\\
 $t^{\text{IMB}}_{\text{off}}$ [s]  &$0.0 ^{+0.13}_{-0}$ &$0.0 ^{+0.13}_{-0}$ &$0.0 ^{+0.13}_{-0}$ & 0-6\\
  $t^{\text{Bak}}_{\text{off}}$ [s] &$0.0 ^{+0.41}_{-0}$ &$0.0 ^{+0.42}_{-0}$ &$0.0 ^{+0.41}_{-0}$ & 0-6\\
 \hline
\end{tabular}
\end{table}


\begin{table}[ht!]
\centering
\caption{Range and best-fit (BF) for all parameters in the time-dependent model Cooling+Accretion. We show the best-fit for three flavor conversion scenarios: MSW with NH, MSW with IH, and a model-independent free $P_{\overline{e}\overline{e}}$.}
\label{tab:parameters_time-dependent}
\begin{tabular}{ lllll } 
 \hline
Parameter & NH BF & IH BF & Free $P_{\overline{e}\overline{e}}$ BF & Range \\ 
 \hline
 $T_{0,c}$ [MeV]   & $4.70^{+0.70}_{-0.60}$ & $4.0^{+0.6}_{-1.5}$ & $5.2^{+0.8}_{-0.8}$ & 1-10 \\
 $\tau_c$ [s]      & $4.5^{+1.3}_{-1.0}$    & $4.8^{+1.6}_{-1.1}$       & $4.5^{+1.3}_{-1.0}$ & 1-40 \\
 $R_c$ [km]        & $14.0^{+7.0}_{-5.0}$   & $15^{+8.0}_{-5.0}$      & $13^{+6.0}_{-4.0}$ & 1-100\\
 $T_{0,a}$ [MeV]   & $1.95^{+0.19}_{-0.13}$ & $3.15^{+0.20}_{-0.18}$    & $1.86^{+0.20}_{-0.12}$ & 0.1-10\\
 $\tau_a$ [s]      & $0.60^{+0.37}_{-0.21}$ & $0.66^{+2.8}_{-0.35}$     & $0.60^{+0.35}_{-0.21}$ & 0.3-3.5\\
 $M_a$ [$M_\odot$] & $0.60^{+0}_{-0.44}$    & $0.6^{+0}_{-0.16}$        & $0.6^{+0}_{-0}$ & 0-0.6\\
 
 $t^{\text{KII}}_{\text{off}}$ [s]  & $0.0^{+0}_{-0}$  & $0.0^{+0.046}_{-0}$ & $0.0^{+0.032}_{-0}$ & 0-6\\
 
 $t^{\text{IMB}}_{\text{off}}$ [s]  & $0.5^{+0.4}_{-0.5}$  & $0.0^{+0.077}_{-0}$ & $0.50^{+0.43}_{-0.31}$ & 0-6\\
 
  $t^{\text{Bak}}_{\text{\text{off}}}$ [s] & $0.0^{+0}_{-0}$   & $0.0^{+0.11}_{-0}$ & $0.0^{+0.10}_{-0}$ & 0-6\\
 \hline
\end{tabular}
\end{table}



\subsection{Time-Integrated}

\begin{figure*}[h!]
    \centering
    \includegraphics[width=\linewidth]{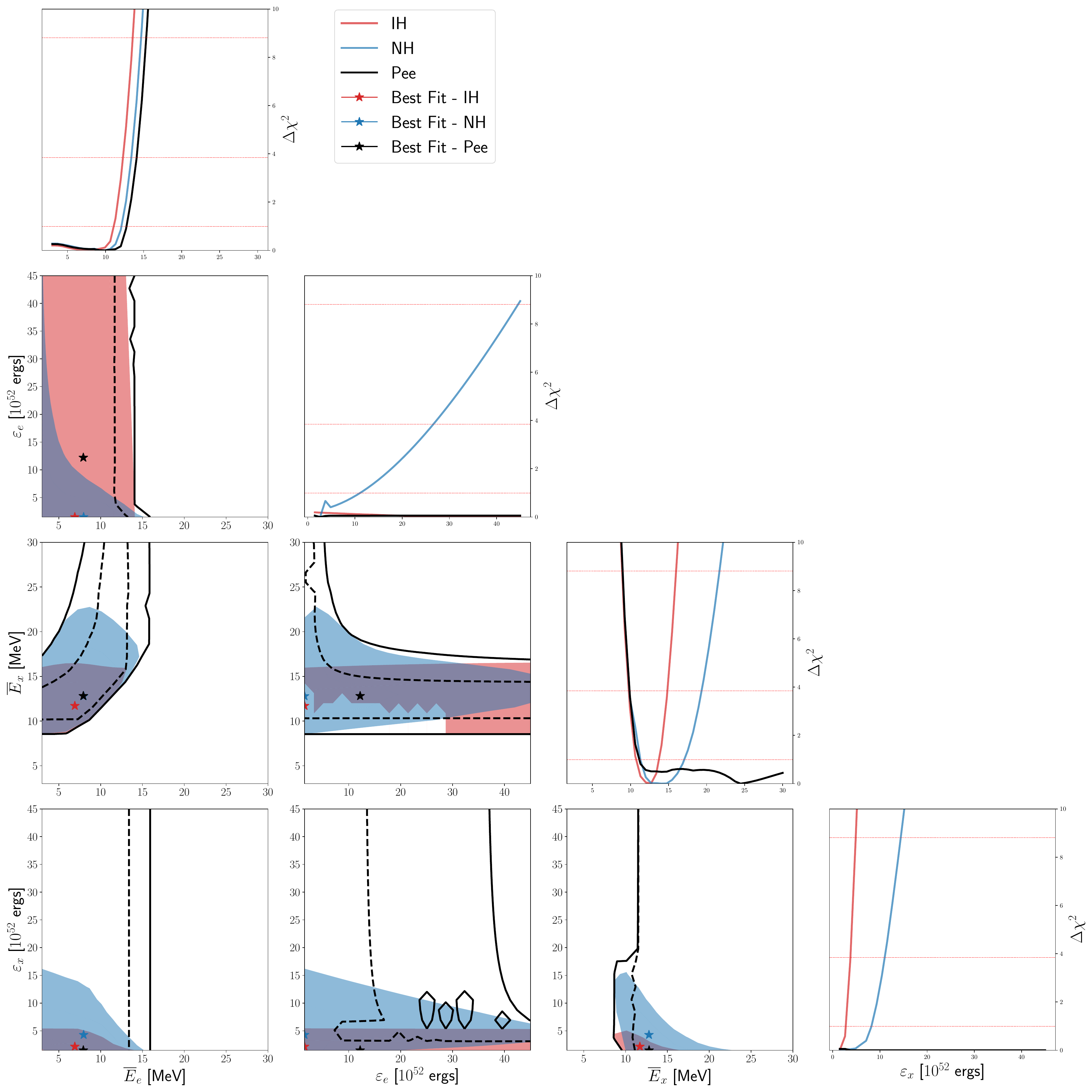}
      \caption{Same as figure \ref{fig:Param_Triangle_Profile_Time_Dependent_Coo}, but for the time-integrated model with astrophysical parameters $\bar{E}_e,\varepsilon_e,\bar{E}_x,\varepsilon_x$.}
      \label{fig:Param_Triangle_Profile_Time_Integrated}
\end{figure*}

For the time-integrated model, we use the work of C. Lunardini \cite{Lunardini:2005jf} for comparison. Figure \ref{fig:Param_Triangle_Profile_Time_Integrated} shows the marginalized profile and contour plots for all the four parameters $\bar{E}_e,\varepsilon_e,\bar{E}_x,\varepsilon_x$ for the three flavor conversion scenario. As already discussed in the paper, there is a preference for $\phi_{\bar{\nu}_e} \approx \phi^0_{\bar{\nu}_x}$, with almost no bound on $\varepsilon_e$ and only a hard upper bound in $\bar{E}_e$ due to the imposed hierarchy in the mean energy. This is consistent with the results shown in Table 1 of \cite{Lunardini:2005jf}.


\begin{table}[ht!]
\centering
\caption{Range and best-fit (BF) for all parameters in the time-integrated model. We show the best-fit for three flavor conversion scenarios: MSW with NH, MSW with IH, and a model-independent free $P_{\overline{e}\overline{e}}$.}
\label{tab:parameters_time-integrated}
\begin{tabular}{ llllll } 
 \hline
Parameter & NH BF & IH BF & Free $P_{\overline{e}\overline{e}}$ BF & Range \\ 
 &  &  &  & \cite{Lunardini:2005jf}\\ 
 \hline
 $\bar{E}_e\; [\text{MeV}]$ & $8.0^{+5.0}_{-5.0}$ & $7^{+5}_{-4}$ & $8.0^{+4.0}_{-5.0}$  & $3-30$   \\
 
 $\varepsilon_e\; [10^{52} \text{ergs}]$ & $1.5^{+10.1}_{-0}$ & $1.5^{+37.2}_{-0}$   & $12^{+33}_{-12}$  & $1.5-45$ \\
 
 $\bar{E}_x\; [\text{MeV}]$ & $12.8^{+1.8}_{-2.0}$ & $11.7^{+1.2}_{-1.1}$ & $12.8^{+1.9}_{-2.1}$  & $3-30$   \\
 
 $\varepsilon_x\; [10^{52} \text{ergs}]$ & $4.3^{+4.0}_{-2.8}$ & $2.2^{+0.8}_{-0.7}$     & $1.5^{+41.5}_{-0}$  & $1.5-45$ \\
 \hline
\end{tabular}
\end{table}

A more direct comparison can be done with the contour plots of $\bar E_{e} \times \bar E_{x}$ and $\bar E_{x} \times \varepsilon_{x}$, which are explicitly shown in Figure 3 of \cite{Lunardini:2005jf}. Our obtained bounds are similar to the one from \cite{Lunardini:2005jf}, where we get stronger bounds in $\bar{E}_x$ in the flavor conversion with fixed $P_{\overline{e}\overline{e}}$, i.e., the MSW scenario with fixed mass hierarchy (NH or IH). This is expected given that $\sin \theta_{13}$ is treated as a free parameter in \cite{Lunardini:2005jf}, which results in a free $P_{\overline{e}\overline{e}}$ within a specif range\footnote{The range used in \cite{Lunardini:2005jf} correspond to the interval $10^{-7}<\sin^2 \theta_{13} <10^{-2}$, which is smaller than our range $[0,1]$.}, given a bound similar to our free $P_{\overline{e}\overline{e}} \in [0,1]$. A similar behavior is found for the bounds on the $\bar E_{x} \times \varepsilon_{x}$ contour plot, where the results of \cite{Lunardini:2005jf} are somewhere between our fixed (NH or IH) and free $P_{\overline{e}\overline{e}}$ scenarios, where in the last scenario no bound is found for $\varepsilon_e$. With this picture in mind, we can conclude that our analysis of the time-integrated model is in relatively good agreement with previous works, given the peculiarities discussed above.



\section{Detection information}
\label{app:Detection}

\begin{figure}=
    \centering
    \includegraphics[width=0.4\textwidth]{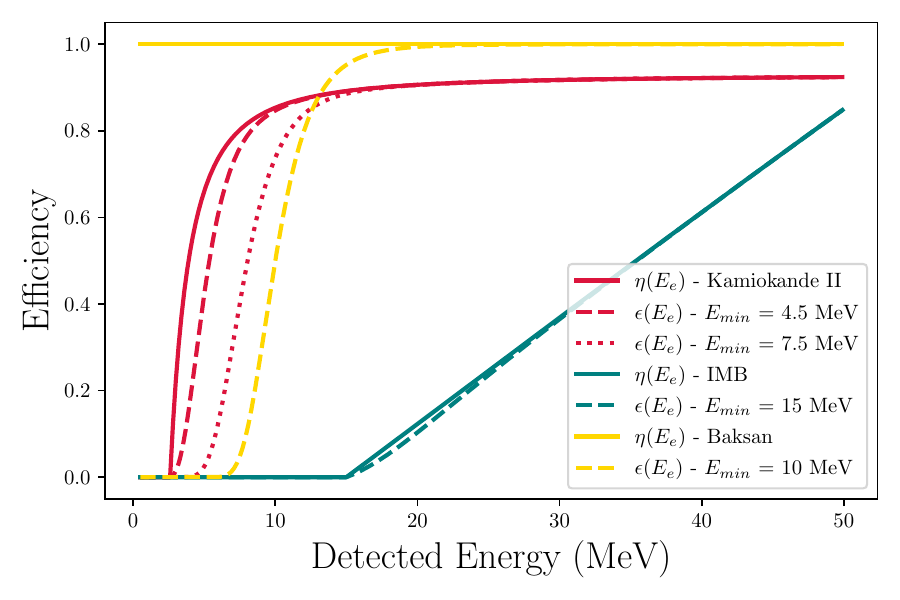}
    \caption{Intrinsic (solid curves) and total efficiencies (dashed and dotted curves) for each detector.}
    \label{fig:Detec_Eff}
\end{figure}

In this appendix, the reader can found information about the detection properties and data used in this work. In Table \ref{tab:Detectors} we show the detectors properties and in Figure \ref{fig:Detec_Eff} the considered efficiency function. By last, we show the neutrino data form Kamiokande-II, IMB, and Baksan in Tables \ref{tab:Kamiokande_data}, \ref{tab:IMB_data}, and \ref{tab:Bakasan_data} respectively.

\begin{table}
\centering
\caption{Characteristics of each detector}
\label{tab:Detectors}
\begin{tabular}{llll} 
 \hline
 Detector & Kamiokande-II & IMB    & Baksan\\ 
 \hline
 Fiducial Mass [kton] & 2.14 & 6.80 & 0.20 \\
 Free Protons [$10^{32}$] & 1.43 & 4.54 & 0.19 \\
 Composition & $H_2O$ & $H_2O$ & $C_9H_2O$ \\
 $\sigma_{stat}$ [MeV] & 1.27 & 3.0 & 0.0 \\
 $\sigma_{syst}$ [MeV ]& 1.00 & 0.4 & 2.0 \\
 \hline
\end{tabular}
\end{table}

\begin{table}
\caption{SN1987A data from Kamiokande-II.}
\label{tab:Kamiokande_data}
\begin{tabular}{lllll}
\hline
\multicolumn{5}{c}{\textbf{Kamiokande-II}} \\
Event & Time & Energy & Angle & Background \\
 & [s] & [MeV] & [Degree] & [MeV$^{-1}$.s$^{-1}$]\\\hline
 
1      & 0         & 20$\pm$2,9             & 18$\pm$18    &$1.0 \times 10^{-5}$  \\
2      & 0,107     & 13,5$\pm$3,2             & 40$\pm$ 27 & $5.4 \times 10^{-4}$ \\
3      & 0,303     & 7,5$\pm$               & 108$\pm$32   & $2.4 \times 10^{-2}$  \\
4      & 0,324     & 9,2$\pm$2,7             & 70$\pm$30   & $2.8 \times 10^{-3}$  \\
5      & 0,507     & 12,8$\pm$2,9             & 135$\pm$23 & $5.3 \times 10^{-4}$  \\
6      & 0,686     & 6,3$\pm$1,7             & 68$\pm$77   & $7.9 \times 10^{-2}$ \\
7      & 1,541     & 35,4$\pm$8               & 32$\pm$16  & $5.0 \times 10^{-6}$   \\
8      & 1,728     & 21$\pm$4,2             & 30$\pm$18    & $1.0 \times 10^{-5}$  \\
9      & 1,915     & 19,8$\pm$3,2             & 38$\pm$22  & $1.0 \times 10^{-5}$  \\
10     & 9,219     & 8,6$\pm$2,7             & 122$\pm$30  & $4.2 \times 10^{-3}$  \\
11     & 10,433    & 13$\pm$2,6             & 49$\pm$26    & $4.0 \times 10^{-4}$  \\
12     & 12,439    & 8,9$\pm$1,9             & 91$\pm$39   & $3.2 \times 10^{-3}$  \\
13     & 17,641    & 6,5 $\pm$1,6             & ---        & $7.3 \times 10^{-2}$  \\
14     & 20,257    & 5,4$\pm$1,4             & ---         & $5.3 \times 10^{-2}$  \\
15     & 21,355    & 4,6$\pm$ 1,3             & ---        & $1.8 \times 10^{-2}$ \\
16     & 23,814    & 6,5$\pm$1,6             & ---         & $7.3 \times 10^{-2}$               \\\hline
\end{tabular}
\end{table}

\begin{table}[h]
\caption{SN1987A data from IMB.}
\label{tab:IMB_data}
\begin{tabular}{lllll}
\hline
\multicolumn{5}{c}{\textbf{\text{IMB}}} \\
Event & Time & Energy & Angle & Background \\
 & [s] & [MeV] & [Degree] & [MeV$^{-1}$.s$^{-1}$]\\\hline
1      & 0         & 38$\pm$7               & 80$\pm$10     &  0            \\
2      & 0,412     & 37$\pm$7               & 44$\pm$15     &  0          \\
3      & 0,65      & 28$\pm$6               & 56 $\pm$20    &  0          \\
4      & 1,141     & 39$\pm$7               & 65$\pm$20     &  0          \\
5      & 1,562     & 36$\pm$9               & 33$\pm$5      &  0         \\
6      & 2,684     & 36$\pm$6               & 52$\pm$0      &  0       \\
7      & 5,01      & 19$\pm$5               & 42$\pm$20     &  0         \\
8      & 5,582     & 22$\pm$5               & 104$\pm$20    &  0         \\\hline
\end{tabular}
\end{table}

\begin{table}[h]
\caption{SN1987A data from Baksan.}
\label{tab:Bakasan_data}
\begin{tabular}{lllll}
\hline
\multicolumn{5}{c}{\textbf{Baksan}}\\
Event & Time & Energy & Angle & Background\\
 & [s] & [MeV] & [Degree] & [MeV$^{-1}$.s$^{-1}$]\\\hline
1      & 0         & 12$\pm$2,4               & ---            &$8.4 \times 10^{-4}$  \\
2      & 0,435     & 17,9$\pm$3,6             & ---            &$1.3 \times 10^{-3}$  \\
3      & 1,71      & 23,5$\pm$4,7             & ---            &$1.2 \times 10^{-3}$  \\
4      & 7,687     & 17,6$\pm$3,5             & ---            &$1.3 \times 10^{-3}$  \\
5      & 9,099     & 20,3$\pm$4,1             & ---            &$1.3 \times 10^{-3}$                \\\hline
\end{tabular}
\end{table}

\end{document}